\documentclass[twocolumn]{aastex701}

\graphicspath{{./}{figures/}}
\usepackage{pifont}
\usepackage{multirow}
\usepackage{amsmath}
\usepackage{color}
\usepackage{graphicx}
\usepackage{subfigure}
\usepackage{boxedminipage}
\usepackage{threeparttable}
\usepackage{cases}
\usepackage{CJKutf8}
\usepackage{lineno}
\usepackage{amssymb}

\shortauthors{Wang et al.}

\begin{document}
\begin{CJK*}{UTF8}{gbsn}

\title{A Two-Stage Kick Scenario for the Peculiar LMXB GX 1$+$4}

\correspondingauthor{Xiang-Dong Li}
\email{lixd@nju.edu.cn}

\author[0000-0002-9738-1238]{Xiangyu Ivy Wang (王翔煜)}
\affiliation{School of Astronomy and Space Science, Nanjing University, Nanjing 210093, China}
\email{mg20260012@smail.nju.edu.cn}

\author[0000-0002-0822-0337]{Shi-Jie Gao}
\affiliation{School of Astronomy and Space Science, Nanjing University, Nanjing 210093, China}
\email{gaosj@smail.nju.edu.cn}

\author[0000-0002-0584-8145
]{Xiang-Dong Li}
\affiliation{School of Astronomy and Space Science, Nanjing University, Nanjing 210093, China}
\affiliation{Key Laboratory of Modern Astronomy and Astrophysics (Nanjing University), Ministry of Education, China}
\email{lixd@nju.edu.cn}

\begin{abstract}

The low-mass X-ray binary (LMXB) GX 1$+$4 stands out with its unique properties. Despite being an old system, it hosts a strongly magnetized neutron star (NS), a trait usually linked to younger systems. Its exceptionally long orbital period (1160 days) and low eccentricity (0.101) imply that the NS formed with minimal mass loss and a weak natal kick. These features collectively point towards the NS having formed through the accretion-induced collapse (AIC) of a white dwarf (WD).
However, GX 1$+$4's unusually high peculiar velocity ($\sim 189.36$ km\,s$^{-1}$) defies standard AIC explanations. To address this discrepancy, we propose a two-stage kick scenario within the AIC framework: 
an initial natal kick followed by a delayed electromagnetic “rocket effect” kick.
Our Monte Carlo simulations indicate that while the natal kick ($\lesssim 100$ km\,s$^{-1}$) can generate a wide range of orbital eccentricities, the subsequent rocket kick ($\sim 240-480$ km\,s$^{-1}$) explains both the high systemic velocity  and low eccentricity. This two-stage kick mechanism naturally reproduces the observed characteristics of GX 1$+$4, provided that the NS's initially buried magnetic field re-emerges after the acceleration process ends.
Our study represents the first attempt to quantitatively constrain the kick velocities in GX 1$+$4 and underscores the importance of possible rocket kicks in forming such peculiar LMXB systems.
\end{abstract}


\section{Introduction} \label{sec:intro}

Low-mass X-ray binaries (LMXBs) are binary systems consisting of a compact object -- either a neutron star (NS) or a black hole -- that accretes matter from a low-mass ($\lesssim$ 1.5 $M_{\odot}$) companion star \citep[][for a recent review]{Tauris2023book}. The supernova (SN) explosion forming the NS typically results in significant mass loss and  imparts a natal momentum kick to the newborn NS, both of which can disrupt the binary.  Due to their relatively low binding energies, surviving LMXBs  usually exhibit short orbital periods ($P_{\rm orb}$) or have experienced weak natal kicks during the SN \citep{Brandt1995MNRAS,Podsiadlowski2002ApJ}. Most LMXBs have $P_{\rm orb}$ $<$ 100 days, with systems exceeding 1000 days being rare \citep[see Figure 6.8 in][]{Tauris2023book}. To date, only three wide-orbit NS LMXBs with orbital periods exceeding 100 days have been identified: IGR J16194-2810 \citep[$P_{\rm orb} \simeq 193$ days;][]{Hinkle2024ApJ, Nagarajan2024PASP}, 4U 1700$+$24 \citep[$P_{\rm orb} \simeq 4391$ days;][]{Hinkle2019ApJ}, and GX 1$+$4 \citep[$P_{\rm orb} \simeq 1160$ days;][]{Hinkle2006ApJ}.

GX 1$+$4  stands out with a peculiar velocity ($v_{\rm pec}$) of about 189.3 km\,s$^{-1}$, the fourth highest among the known LMXBs \citep{Zhaoyue2023MNRAS}. Its combination of high peculiar velocity, long $P_{\rm orb}$, and low eccentricity \citep[$e$ = 0.101;][]{Hinkle2006ApJ} challenge conventional natal kick models, which cannot simultaneously explain these properties. Furthermore, the NS’s inferred surface magnetic field ($>10^{12}$ G) suggests a relatively young NS, contradicting the evolutionary age of its old red giant (RG) companion. 

Several formation pathways for GX 1$+$4 have been proposed in the literature \citep[see also][]{Hinkle2006ApJ}: 
\begin{enumerate}
    \item[(1)] Capture of an NS in a globular cluster. Successful capture requires the NS to approach within 2–3 times the target star’s radius, a condition mostly achievable in globular cluster cores \citep{Canal1990ARA&A}. However,  such a wide-orbit system would struggle to escape the cluster’s gravitational potential \citep{Belloni2020MNRAS}. Hence, this scenario seems unlikely.
    \item[(2)] Isolated evolution of a primordial binary to NS + RG binary. The primary challenge for this scenario is how to keep the system bound during the primary star’s collapse into an NS \citep[see][for a discussion]{Hinkle2006ApJ}. If the binary survives, the secondary star then evolves off the main sequence (MS) and enters the RG branch or the asymptotic giant branch after $\sim 5\times 10^9$ yr \citep{Charbonnel1996}. This conflicts with the NS’s youth ($\lesssim 10$ Myr, inferred from its strong magnetic field), creating a timescale inconsistency between the young, magnetized NS and its old RG companion.
    \item[(3)] Triple star evolution. \cite{Canal1990ARA&A} and \cite{Iben1999ASPC} proposed a triple system where a massive close binary (with a low-mass MS tertiary) undergoes mass transfer and a SN, leaving a massive star and an NS. The NS and the companion then enter the common-envelope phase, forming a Thorne-Z\'ytkow object (RG with an NS core). After envelope ejection, an NS + MS binary remains.  However, this scenario requires fine-tuned conditions and shares the same timescale issues as Scenario 2. 
    \item[(4)] Accretion-induced collapse (AIC) of a white dwarf (WD) \citep{Heuvel1984}. A massive WD accretes material from an evolved companion star, reaching the Chandrasekhar limit, and collapses into an NS \citep[see][]{Wangbo2020RAA}. The resulting binary contains a young NS and a giant star. Recent studies suggested that the AIC mechanism may efficiently form highly magnetized NS + RG binaries \citep{Ilkiewicz2019MNRAS, Ablimit2023MNRAS}, supporting  its viability as the origin of systems like GX 1$+$4.
\end{enumerate}

While the AIC process  offers a plausible formation pathway for GX 1$+$4, it struggles to account for the system's high peculiar velocity. It is widely acknowledged that NSs born through AIC typically receive only modest natal kicks, owing to the minimal mass loss associated with WD implosion \citep{Fryer1999ApJ, Dessart2006ApJ, Tauris2013A&A}. This suggests that additional mechanism(s) must be at play to explain the substantial peculiar velocity observed in GX 1$+$4.

One potential explanation is the ``rocket effect", which provides supplementary acceleration to the newborn NS.  Unlike natal kicks  that occur during SN explosions, the rocket effect acts post-explosion and unfolds over an extended timescale \citep{Harrison1975ApJ, Lai2001ApJ}.  This effect arises from asymmetric electromagnetic dipole radiation, which imparts a velocity kick along the NS's spin axis and operates on the spin-down timescale. It has been invoked to explain the high velocities and spin-velocity alignment observed in pulsars such as PSR J0538$+$2817 \citep{Xu&Huang2022MNRAS}. \cite{Hirai2024ApJ} explored the combined influence of natal and rocket kicks in shaping wide, low-$e$ NS binaries. They posited that an SN explosion would initially drive the binary into a wide, high-$e$ orbit, with the rocket effect subsequently acting to reduce the eccentricity. However, their study did not address the specific scenario where the NS in GX 1$+$4 formed via WD implosion rather than core-collapse SN. Given that the AIC process generates only small natal kicks, the system may not initially attain a wide, high-$e$ orbit during NS formation. This implies that the rocket effect alone could be the primary driver of GX 1$+$4's high peculiar velocity.

In this study, our objective is to offer a unified explanation for the formation of a highly magnetized young NS paired with an old RG companion, characterized by a long orbital period, low eccentricity, and high peculiar velocity, within the framework of AIC combined with the rocket kick mechanism. By comparing our model with observational data from GX 1$+$4, we aim to derive quantitative constraints on the kick velocities. The structure of this paper is as follows. In Section \ref{sec:observations}, we present the observational properties of GX 1$+$4. In Section \ref{sec:formation_scenarios}, we detail the methodology  employed  to constrain the kick velocities and present our findings. Finally, in Section \ref{sec:conclusions}, we briefly summarize our results and discuss their implications.

\section{Observational Properties of GX 1$+$4} \label{sec:observations}

\begin{figure}
    \centering
    \includegraphics[width=\linewidth]{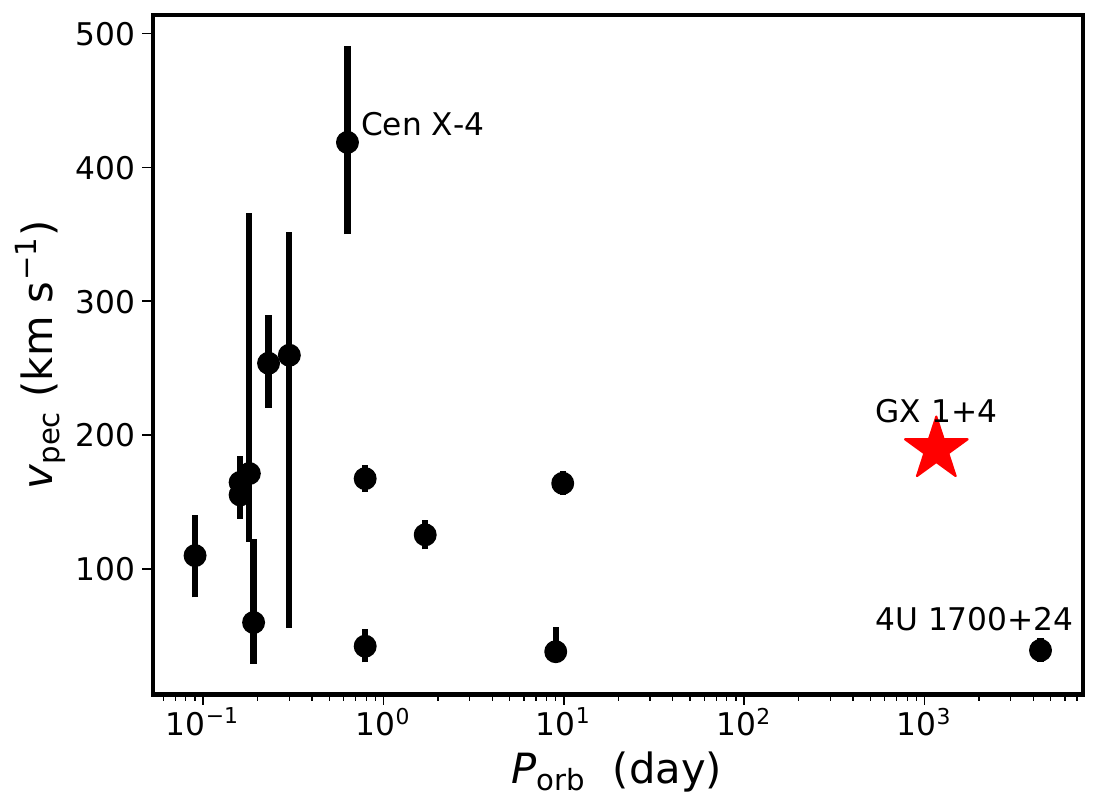}
    \caption{The $v_{\rm pec}$-$P_{\rm orb}$ distribution of LMXBs in \cite{Zhaoyue2023MNRAS}'s catalog. The red star represents GX 1$+$4.}
    \label{fig:v-p}
\end{figure}

\begin{figure}
    \centering
    \includegraphics[width=\linewidth]{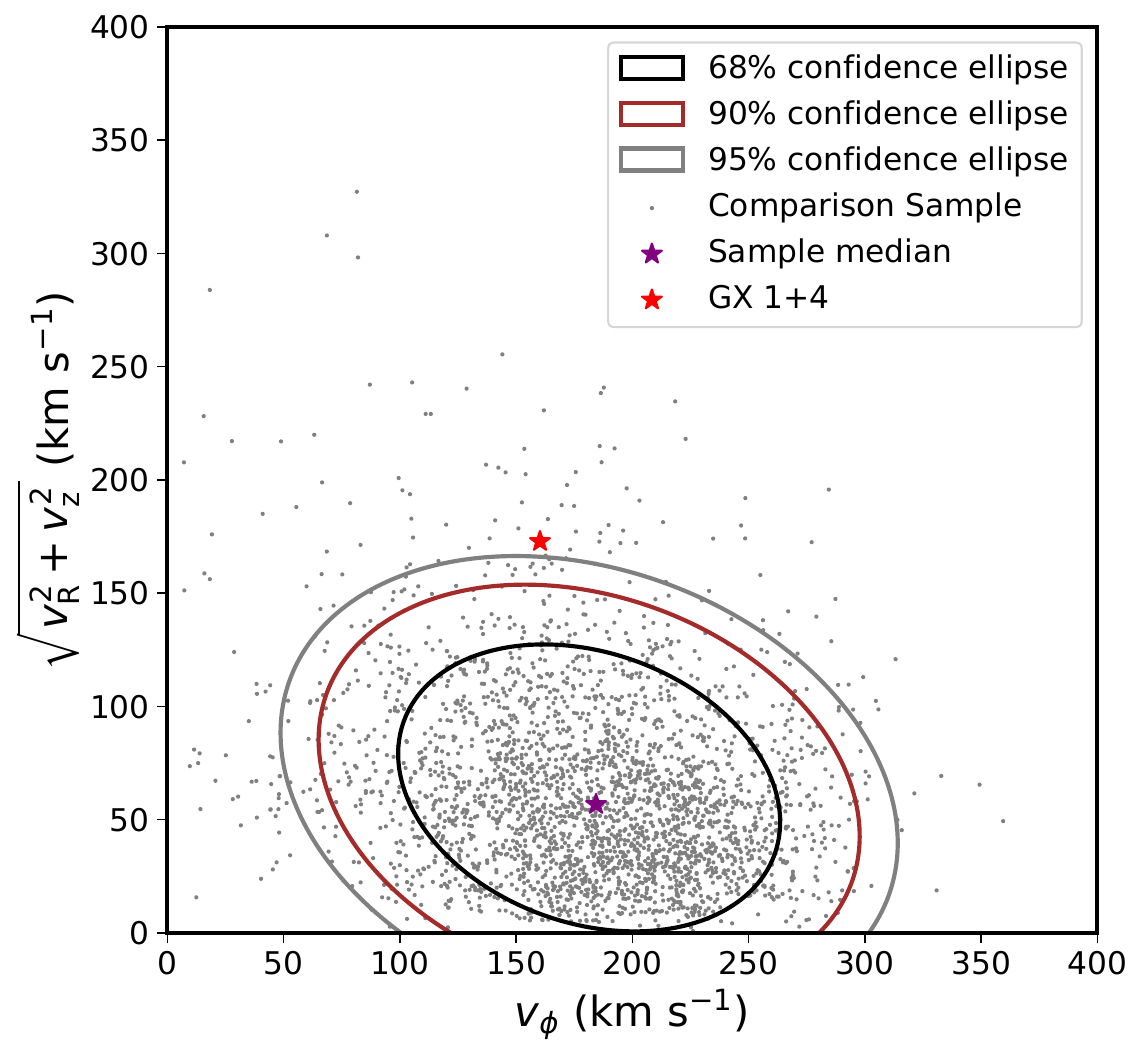}
    \caption{Toomre diagram for GX 1$+$4 and its local comparison sample. Stars from Gaia DR3 are plotted as grey points for comparison. GX 1$+$4 is marked with a red star, while the sample median is indicated by a purple star. The solid black, brown, and grey ellipses enclose 68\%, 90\%, and 95\% of the comparison sources, respectively.}
    \label{fig:Toomre}
\end{figure}

GX 1$+$4 is a well-known symbiotic X-ray binary system consisting of a slowly rotating NS that is accreting the wind material from its RG companion \citep{Lewin1971ApJ, Davidsen1977ApJ, Hinkle2006ApJ}. It is in a 1161-day orbit with an eccentricity of 0.101 \citep{Hinkle2006ApJ}. The NS has undergone a spin reversal, transitioning from a spin-up to a spin-down phase, and currently exhibits a spin period of around 180 s \citep{Ricketts1982MNRAS, Luna2023A&A}. \cite{Gonzalez2012A&A} suggested that the long-term spin-down behaviour could be attributed to quasi-spherical accretion \citep{Shakura2012MNRAS} from the stellar wind.

Several lines of evidence indicate that GX 1$+$4 hosts a strongly magnetized NS.
First, the observed spectrum of GX 1$+$4 is among the hardest detected in all X-ray binary pulsars, which suggests a magnetic field strength ($B$) exceeding $10^{13}$ G \citep{Coburn2002}.
Second, there is marginal evidence for a cyclotron resonance scattering feature at $33-35$ keV, indicating a magnetic field of approximately $3\times10^{12}$ G \citep{Rea2005MNRAS, Ferrigno2007A&A}. Additionally, the possible propeller effect observed in GX 1$+$4 implies an even stronger magnetic field, on the order of $10^{13}-10^{14}$ G \citep{Cui1997ApJ,Cui2004}. Such a strong magnetic field suggests that the NS is relatively young, with an age of no more than a few $10^6$ yr \citep{Aguilera2008A&A, Popov2012MNRAS}. In contrast, the optical companion, V2116 Ophiuchi \citep{Glass1973NPhS, Chakrabarty1997ApJ} is near the tip of the RG branch, with an effective temperature of about 3400 K, a surface gravity of $\log\,g$ (cm\,s$^{-2}$) = 0.5, and does not fill its Roche lobe \citep{Hinkle2006ApJ}. Assuming an NS mass of 1.35 $M_{\odot}$, the companion mass is estimated to be 1.22 $M_{\odot}$ \citep{Hinkle2006ApJ}. The companion's age \citep[$\sim 5\times 10^9$ yr,][]{Charbonnel1996,Hinkle2006ApJ} is significantly greater than that of the NS.  

Infrared observations of the RG companion place the system at a distance of 4.3 kpc, with a systemic radial velocity of $176.73\pm 0.22$ km\,s$^{-1}$ \citep{Hinkle2006ApJ}. Combining these data with proper motion measurements from \textit{Gaia} Data Release 3 \citep[\textit{Gaia} DR3; $\mu_\alpha\cos\delta = -3.52\pm 0.08$ mas\,yr$^{-1}$ and $\mu_\delta = -1.99\pm 0.06$ mas\,yr$^{-1}$;][]{GAIA2023A&A}, the system's peculiar velocity was calculated to be $189.3^{+8.1}_{-8.3}$\,km\,s$^{-1}$ \citep{Zhaoyue2023MNRAS}. 

If the NS in GX 1$+$4 was formed through AIC,  tidal torques would circularize the binary orbit during subsequent evolution. According to \citet{Soker2000A&A} and \citet{Hinkle2006ApJ}, the tidal circularization timescale is estimated to be approximately $5 \times 10^6 \sin^{-8}i$ yr, where $i$ is the inclination angle of the binary orbit. This timescale is significantly longer than the age of the NS, unless $i$ is extremely close to $90\degr$. Thus, we conclude that the binary orbit has not been substantially circularized since the NS's formation. 

Figure \ref{fig:v-p} illustrates the distribution of peculiar velocities and orbital periods for 15 LMXBs from the catalog of \cite{Zhaoyue2023MNRAS}. Among these sources, GX 1$+$4 is unique in exhibiting both a long $P_{\rm orb}$ and a high $v_{\rm pec}$. Sources with  high $v_{\rm pec}$  generally reside in compact orbits ($P_{\rm orb}\lesssim 10$ days). Although 4U 1700$+$24 has a very long $P_{\rm orb}$, the binary possesses a low peculiar velocity \citep[$v_{\rm pec}\sim$39.2 km\,s$^{-1}$;][]{Zhaoyue2023MNRAS}. Furthermore,  high $v_{\rm pec}$ is often associated with high eccentricities (see Section \ref{subsec:kick_simulation} for details), yet GX 1$+$4 maintains a low-eccentric orbit. 

To compare the kinematics of GX 1+4 with the local stellar population, we constructed a Toomre diagram for this source and its comparison sample. The comparison sample was drawn from \texttt{Gaia} DR3, satisfying the following criteria: (1) \texttt{parallax$\_$over$\_$error} $\geq$ 3; (2) radial velocity error $<$ 5 km\,s$^{-1}$; (3) proper motion errors $<$ 1 mas\,yr$^{-1}$ for both components; and (4) located within a 500 pc radius of GX 1+4. The Galactocentric space velocity ($V_{\rm R}$, $V_{\rm z}$, $V_{\phi}$) for all sources were calculated in cylindircal coordinates using \texttt{PYTHON} package \texttt{Astropy} \citep{Astropy2013A&A, Astropy2018AJ}. We adopted a solar distance to the Galactic center of $R_{\odot}$ = 8.122 kpc \citep{GRAVITY2018A&A}, a solar height above the Galactic midplane of $z_{\odot}$ = 20.8 pc \citep{Bennett2019MNRAS}, and a solar velocity of ($V_{\rm R,\odot}$, $V_{\phi,\odot}$, $V_{\rm z, \odot}$) = ($-12.9$, 245.6, 7.78) km\,s$^{-1}$ \citep{Drimmel2018RNAAS}. The resulting Toomre diagram (Figure \ref{fig:Toomre}) plots the non-azimuthal velocity ($\sqrt{V_{\rm R}^2 + V_{\rm Z}^2}$) versus the azimuthal velocity ($V_{\phi}$). The median velocity of the comparison sample is marked with a purple star, while the 68\%, 90\%, and 95\% confidence ellipses are shown in black, brown, and grey, respectively. GX 1+4, indicated by a red star, lies outside the 95th percentile contour of the comparison sample, providing evidence of a strong kick.

The conventional NS kick model, which involves SN explosions, faces difficulties in simultaneously explaining the high $v_{\rm pec}$, long $P_{\rm orb}$, and low $e$ observed in GX 1$+$4. Additionally, the presence of a highly magnetized young NS alongside an old RG companion suggests that AIC, rather than core-collapse SN, is a more plausible formation mechanism. In the following section, we will investigate how a binary system with high-$v_{\rm pec}$, long-$P_{\rm orb}$, and low-$e$ can be formed through the AIC process and constrain the associated kick velocities.

\section{The AIC Formation Scenario} \label{sec:formation_scenarios}

\subsection{Evolution with Natal Kick} \label{subsec:kick_simulation}

\begin{figure}
    \centering
    \includegraphics[width=\linewidth]{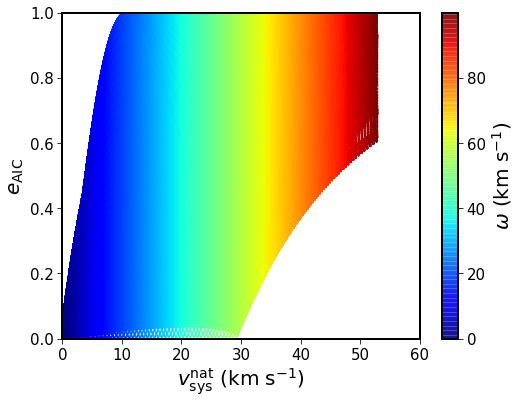}
    \caption{The simulated $e_{\rm AIC}$-$v_{\rm sys}^{\rm nat}$ distribution for various natal kick velocities ($\omega$). }
    \label{fig:fig2}
\end{figure}

In ONeMg WD+RG binaries, mass transfer through stellar wind capture and/or Roche-lobe overflow can drive the WD toward the Chandrasekhar limit, potentially  triggering AIC and forming an NS under specific conditions \citep{Ilkiewicz2019MNRAS, Ablimit2023MNRAS}. The nascent NS is expected to  retain rapid rotation from its WD progenitor, as long-term accretion can significantly  spin up the pre-AIC WD \citep{Yoon2005A&A, Kuroda2025arXiv}. During the collapse, angular momentum conservation  results in a spin period of a few milliseconds for the newborn NS. Numerical simulations indicate that the NS might further acquire strong magnetic fields  up to $\sim 10^{15}$ G \citep{Dessart2007ApJ}, positioning them as potential progenitors of millisecond magnetars \citep{Usov1992Natur}. These magnetars are theorized to power luminous transient events \citep{Yu2015ApJ, Zhu2021ApJ} and emit distinct X-ray/radio signals \citep{Piro2013ApJ, Yu2019ApJ}. Unlike core-collapse SNe,  NSs produced via AIC typically experience minimal kicks, which could be characterized by a Maxwellian distribution with $\sigma_{\rm k}\simeq 50$ km s$^{-1}$ \citep{Hurley2010MNRAS, Tauris2023book}.

We employ the Monte Carlo method to simulate the post-AIC orbit, assuming a circular pre-AIC orbit \citep{Mikolajewska2012BaltA}. We set the masses of the WD and the RG at the AIC to be $M_{\rm WD}=1.4\,M_{\odot}$ \citep[Chandrasekhar limit;][]{Chandrasekhar1931ApJ} and $M_{\rm RG}=1.22\, M_{\odot}$ respectively, and the NS mass $M_{\rm NS}=1.35\,M_{\odot}$. The companion mass remains unchanged  following the AIC. Given that the long-term rocket effect does not affect the binary's orbital period (refer to Section \ref{subsec:rocket} for details), we set the post-AIC orbital period at 1160 d.

The system's total energy change alters the orbital semi-major axis. The ratio of the final to initial orbital semi-major axis is given by 
\citep{Hills1983ApJ, Tauris2017ApJ}:
\begin{equation}
    \label{eq:af_ai}
    \frac{a_{\rm f}}{a_{\rm i}} = \left [ \frac{1 - \Delta M/M}{1 - 2\Delta M/M - (\omega/v_{\rm rel})^2 - 2 \cos \theta (\omega/v_{\rm rel})} \right],
\end{equation}
where $\Delta M = M_{\rm WD}-M_{\rm NS}$ represents mass loss during AIC, $M = M_{\rm WD}+M_{\rm NS}$ is the pre-AIC total mass, $\omega$ is the natal kick velocity, and $v_{\rm rel} = \sqrt{G(M_{\rm WD} + M_{\rm RG})/a_{\rm i}}$ is the pre-AIC relative orbital velocity with $G$ the gravitational constant. The kick orientation is defined by two angles: $\theta$, the angle between $\omega$ and the pre-AIC orbital velocity vector of the WD, and $\phi$, the angle between the projection of $\omega$ on the plane perpendicular to the WD's orbital velocity vector and the orbital plane \citep[See Figure 13 in][for geometric configuration]{Tauris2017ApJ}.

The eccentricity of the post-AIC binary can be calculated using
\begin{equation}
\label{eq:post_e}
    e_{\rm AIC} = \sqrt{1 + \frac{2 E_{\rm orb,f}  L_{\rm orb,f}^2}{\mu_{\rm f}  G^2  M_{\rm NS}^2  M_{\rm RG}^2}},
\end{equation}
where $L_{\rm orb,f}$ is the post-AIC orbital angular momentum, expressed as:
\begin{equation}
\label{eq:Lorb_f}
    L_{\rm orb,f} = a_{\rm i} \mu_{\rm f} \sqrt{(v_{\rm rel}+\omega \cos \theta)^2 + (\omega \sin \theta \sin \phi)^2}.
\end{equation}
Here, $\mu_{\rm f} \equiv M_{\rm NS} M_{\rm RG}/(M_{\rm NS}+M_{\rm RG})$ and $E_{\rm orb,f} = G M_{\rm NS} M_{\rm WD}/2 a_{\rm f}$ are the post-AIC reduced mass and the orbital energy, respectively.

We simulate the AIC process across a three-dimensional parameter space ($\theta$, $\phi$, and $\omega/v_{\rm rel}$), sampling 10,000 isotropic kick directions for $\theta$ and $\phi$, and 1000 uniformly distributed $\omega/v_{\rm rel}$ values in the range of $(0, 2.8)$. From these simulations, we select systems with natal kicks $\omega \leq 100$ km\,s$^{-1}$. The resulting  distribution of $e_{\rm AIC}$ versus $v_{\rm sys}^{\rm nat}$ for different $\omega$ values is shown in Figure \ref{fig:fig2}.

Our simulations indicate that systems with $\omega \leq$ 100 km\,s$^{-1}$ achieve a maximum $v_{\rm sys}^{\rm nat}$ of 52.89 km\,s$^{-1}$, substantially lower than the observed peculiar velocity ($v_{\rm pec}\sim 190$ km\,s$^{-1}$) of GX 1$+$4. When  restricting the analysis to cases with  $\omega \leq$ 50 km\,s$^{-1}$, the maximum $v_{\rm sys}^{\rm nat}$  decreases to 26.46 km\,s$^{-1}$. These numbers suggest that natal kicks contribute minimally to the system's peculiar velocity during the AIC process.

\subsection{Evolution with Rocket Kick} \label{subsec:rocket} 

\begin{figure}
    \centering
    \includegraphics[width=\linewidth]{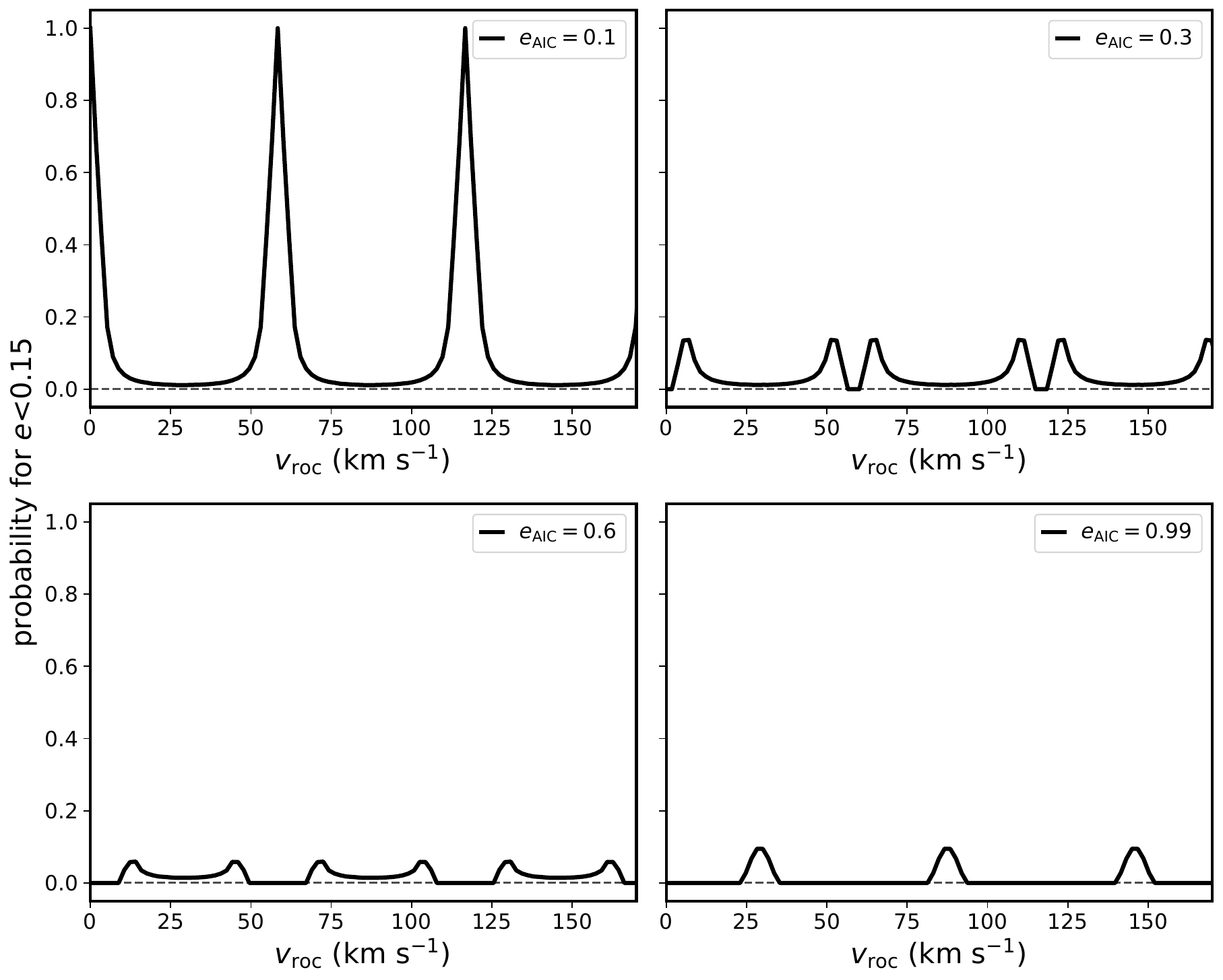}
    \caption{Probability distribution of systems achieving an eccentricity $e<0.15$ as a function of $v_{\rm roc}$ derived from rocket kick simulations, assuming a post-AIC eccentricity of $e_{\rm AIC}$ = 0.1, 0.3, 0.6, and 0.99.}
    \label{fig:fig3}
\end{figure}

\begin{figure}
    \centering
    \includegraphics[width=\linewidth]{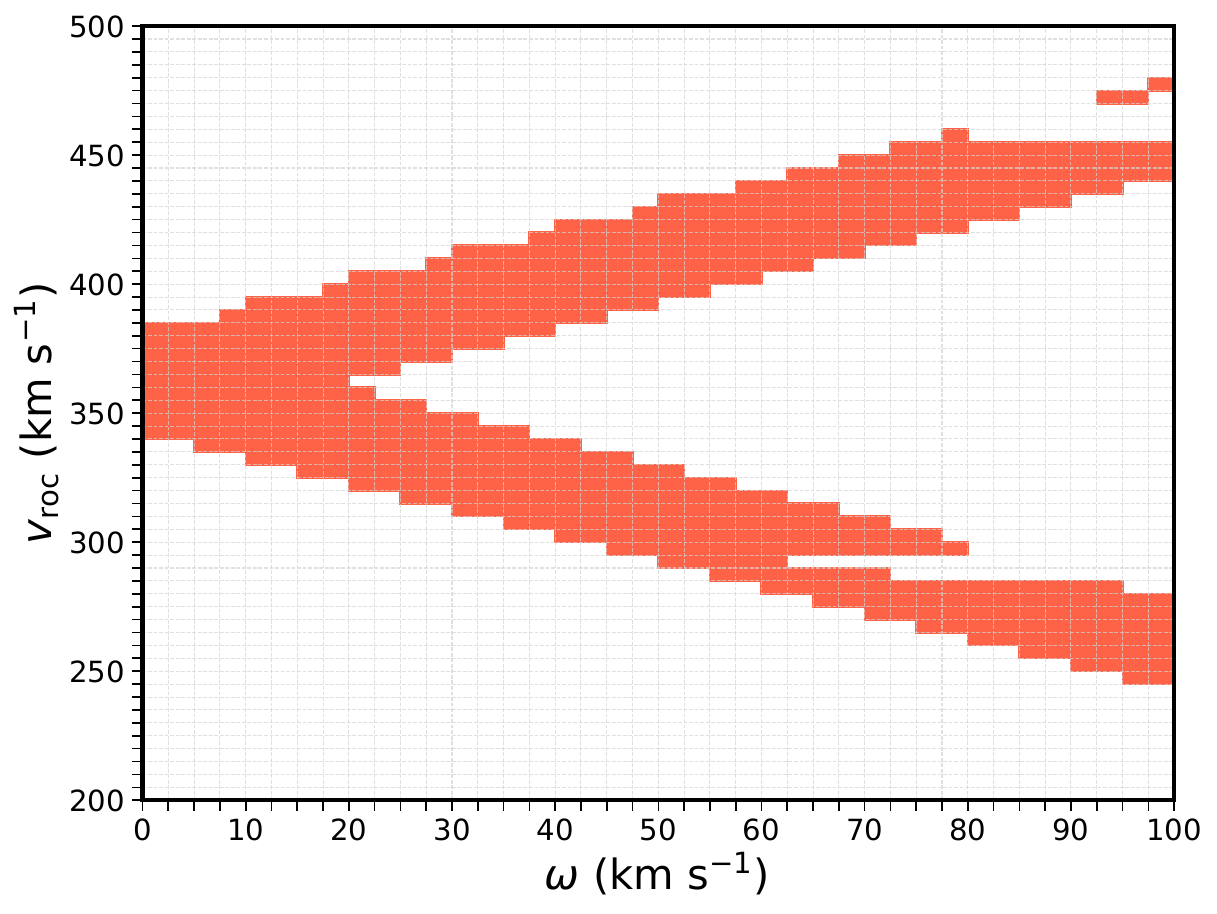}
    \caption{Allowed  $\omega$-$v_{\rm roc}$ parameter space for forming systems like GX 1$+$4.}
    \label{fig:fig4}
\end{figure}

Magnetized NSs are conventionally assumed to have magnetic dipole fields aligned such that the magnetic axis passes through the NS center. However, the magnetic axis may be off-centered. In this case, rotating magnetic fields will generate asymmetric radiation, exerting a force on the NS, which is parallel to the spin axis. This is called the electromagnetic rocket effect \citep{Harrison1975ApJ, Lai2001ApJ, Kojima2011ApJ, Petri2019MNRAS}. The effect has been proposed to explain the high velocities and spin–velocity alignment phenomenon observed in  pulsars \citep{Johnston2007,Noutsos2012,Noutsos2013,Xu&Huang2022MNRAS,Biryukov2025}.

For a dipole moment $\mu$ (with $\mu_{\rm z}$, $\mu_{\phi}$, $\mu_{\rho}$ being its three components in spherical coordinates) displaced by $s_{\rho}$
 from the star's rotation axis，the rocket kick magnitude in vacuum environment is given by \citep{Lai2001ApJ,Baibhav2025arXiv}
\begin{equation}
\label{eq:rocv_vacuum}
\hspace{-1em}
    v_{\rm roc}^{\rm vac} \simeq 180 \ \left( \frac{R_{0}}{12 \ \rm{km}} \right)^2 \left(\frac{s_{\rho}}{12 \ \rm{km}}\right) \left(\frac{\mu_{\rm z}\mu_{\phi}}{\mu_{\phi}^2+\mu_{\rho}^2}\right)\left(\frac{\nu_{0}}{716 \ \rm{Hz}}\right) \rm{km \, s}^{-1},
\end{equation}
where $R_{0}$ is the NS radius, and $\nu_{0}$ is the initial spin frequency. This equation demonstrates that the kick velocity is independent of the magnetic field strength, depending solely on its configuration. For initial spin frequencies $\nu_{0} \simeq 1000$ Hz, the maximum rocket kick can reach approximately 1400 km\,s$^{-1}$ \citep{Lai2001ApJ}. The duration of this acceleration is governed by the NS's spin-down timescale:
\begin{equation}
\label{eq:roc_time}
\hspace{-1em}
    \tau \simeq 30 \left( \frac{I}{10^{45} \,\rm{g\,cm^2}} \right) \left(\frac{B_{0}}{10^{12}  \,\rm{G}}\right)^{-2} \left( \frac{R_{0}}{12\,\rm{km}} \right)^{-6}  \left ( \frac{\nu_{0}}{716 \, \rm{Hz}} \right)^{-2} \rm{yr},
\end{equation}
where $I$ is the moment of inertia and $B_{0}$ is the initial magnetic field. Beyond this timescale, the rocket force becomes negligible, marking the end of the acceleration process.

In binary systems, the rocket effect influences both the systemic velocity and orbital eccentricity \citep{Hirai2024ApJ}.  Although its long-term impact on the orbital period is negligible, it induces periodic variations in eccentricity for fixed kick angles \citep{Petri2019MNRAS, Hirai2024ApJ}. Our simulations reveal that the post-AIC systems  display a wide spectrum of eccentricities ($e_{\rm AIC} \in$ [0,1]), as demonstrated in Figure \ref{fig:fig2}. To explore the impact of rocket kicks on these systems, we simulate the orbit with $e_{\rm AIC}$=(0.1, 0.3, 0.6, 0.99). For each rocket kick velocity vector, we take into account 10,000 potentially isotropic directions, and calculate the probability of achieving a final eccentricity $e<0.15$, determined by the ratio of rocket kick directions that result in $e<0.15$ to the total number of simulated directions. Figure \ref{fig:fig3}  presents the results in the four cases and reveals three key features: (1) The probability displays a 58.36 km\,s$^{-1}$ periodicity in the rocket kick velocity space; (2) When $e_{\rm AIC}$ = 0.1,  all kick velocities can produce a final $e<0.15$ through certain directions, whereas velocities satisfying $v_{\rm roc}\approx 58.36\cdot n$ km\,s$^{-1}$ (where $n$ is an integer) ensure $e<0.15$
for all possible kick directions;
(3) As $e_{\rm AIC}$ increases, the range of $v_{\rm roc}$ values that allow for a probability of  $e < 0.15$ narrows, and the minimum required velocity rises from zero at $e_{\rm AIC}=0.1$ to 24.76 km\,s$^{-1}$ at $e_{\rm AIC}=0.99$.

To identify the kick parameters that can reproduce the observed properties of GX 1$+$4 ($v_{\rm pec}\simeq 190$ km s$^{-1}$, $e\simeq 0.1$), we explore the natal kick and rocket kick velocities in a two-dimensional parameter space. We first divide the natal kick velocity range ($0-100$ km s$^{-1}$) into 50 bins, each 2 km\,s$^{-1}$ wide. Given that the maximum and minimum peculiar velocities within each bin differ by less than 2.5 km\,s$^{-1}$, we use the average systemic velocity per bin as the representative systemic velocity ($v_{\rm sys}^{\rm nat}$) for that natal kick velocity range. We then model the rocket kick velocity as uniformly distribution within $0–1000$  km\,s$^{-1}$ (sampling 200 values) with an isotropic direction distribution (5000 orientations). Finally, for each ($\omega$, $v_{\rm roc}$) bin we evaluate the combined effect by calculating the rocket-induced velocity component $v_{\rm sys}^{\rm roc} = M_{\rm c} v_{\rm roc} / (M_{\rm NS} + M_{\rm c})$ and deriving the total systemic velocity $v_{\rm sys}^{\rm total}$ through vector addition of $v_{\rm sys}^{\rm nat}$ and $v_{\rm sys}^{\rm roc}$. We identify viable configurations where the resulting system achieves $e \in [0.09, 0.11]$ and $v_{\rm sys}^{\rm total} \in [180, 200]$ km\,s$^{-1}$. These configurations require specific angles between the kick velocity vectors. 

Figure \ref{fig:fig4} illustrates the derived $\omega$-$v_{\rm roc}$ distribution that could produce systems like GX 1$+$4. Our analysis indicates that when $\omega$ from the AIC process is within the $0$-$100$ km\,s$^{-1}$ range, the corresponding $v_{\rm roc}$ must lie within $(240, 480)$ km\,s$^{-1}$. Notably, systems with properties akin to GX 1$+$4 can only form when the combined kicks satisfy $340\,\rm{km\,s^{-1}} < $ $v_{\rm roc} \pm \omega \, < 380\,\rm{km\,s^{-1}}$, representing necessary but not sufficient conditions for such systems. 

Recently \citet{Baibhav2025arXiv} argued that the vacuum magnetosphere assumption is invalid in realistic astrophysical conditions. In a force-free magnetosphere, the rocket force vanishes because the electric field parallel to the magnetic field in the rotating frame is screened by pair production, which fills the magnetosphere with plasma. Consequently, substantial kicks can only develop in pulsars with insufficient pair production at the light cylinder -- namely, those with initially rapid spins ($\gtrsim 500$ Hz), weak dipole fields ($B_0\lesssim 10^{10}$ G), and relatively strong quadrupole  components. 
The inferred magnetic field strength of GX 1$+$4 is $10^{12}$-$10^{14}$ G. Such strong fields drive efficient pair production, creating a nearly force-free magnetosphere where the rocket kick would be negligible.

One possible explanation is that the NS’s initial strong magnetic field was buried by material accreted from a fallback disk \citep{Muslimov1995,Geppert1999}.  Radiation-hydrodynamics simulations of AIC of white dwarfs indicate the formation of a post-collapse, quasi-Keplerian $0.1-0.5\,M_\odot$ accretion
disk, suggesting prolonged accretion onto the newborn NS \citep{Dessart2006ApJ}. We propose a scenario for GX 1$+$4 where the NS was born with a strong magnetic field, subsequently buried by fallback accretion, and later re-emerged from the accreted crust \citep[see also][]{BaishengLiu2019RAA, Igoshev2021Univ}. The burial temporarily suppressed pair production, creating a vacuum-like environment that enables highly asymmetric magnetic dipole radiation and generated a significant rocket kick. The buried magnetic field re-emerged via Ohmic dissipation and Hall drift on timescales of $10^4-10^6$ yr \citep{Geppert1999,Ho2011MNRAS,Vigano2012MNRAS, Igoshev2016MNRAS}. Once the field resurfaced, the system returned to a force-free state, halting acceleration. For an NS with an initial spin period of 2 ms and a post-burial magnetic field of $10^{10}$ G, Equation (\ref{eq:roc_time}) predicts an acceleration timescale of a few $10^5$ yr, comparable to the magnetic re-emergence timescale.

\section{Conclusions and Discussion} \label{sec:conclusions}

In this study, we introduce a two-stage formation scenario for the peculiar LMXB GX 1$+$4, which accounts for its wide orbit, high peculiar velocity, and low eccentricity. The process begins with an AIC, resulting in a rapidly rotating NS that experiences a weak natal kick  ($\omega \lesssim 100$ km\,s$^{-1}$). Subsequently, the electromagnetic rocket effect imparts an additional kick velocity ranging from  240 to 480\,km\,s$^{-1}$. This delayed kick not only generates a high systemic velocity but also evolves the system's eccentricity.  Our analysis indicates that a necessary, though not sufficient, condition for replicating the observed properties of GX 1$+$4 is that the combined kick velocities satisfy $340\,\rm{km\,s^{-1}}$$<v_{\rm roc} \pm \omega < 380\,\rm{km\,s^{-1}}$. 

A key implication of our model is that the initially strong magnetic field of GX 1$+$4 was likely buried by fallback accretion following its AIC. This burial process would temporarily create a vacuum-like environment around the NS, enabling asymmetric magnetic dipole radiation to generate a rocket kick comparable to our simulation results. The magnetic field would subsequently re-emerge after the acceleration phase concludes. Our calculations demonstrate that this entire sequence is feasible under specific conditions, supporting the physical plausibility of our proposed scenario.

In contrast to the scenario proposed by \cite{Hirai2024ApJ}, which considers core-collapse SNe where strong natal kicks ($\omega \gtrsim 100$ km\,s$^{-1}$)  initially create wide, highly eccentric orbits that are later circularized by rocket effects, our model offers a different perspective: (1) the collapse induced by AIC results in small natal kicks, preventing the formation of high systemic velocities and leaving the post-AIC eccentricity ranging from 0 to 1; (2) the rocket effect then becomes the primary driver of the system's high peculiar velocity and adjusts the eccentricity to approximately 0.101. Our model provides quantitative constraints on both kick velocities and the initial spin period of the NS, offering new insights into determining the birth properties of NSs.

The main limitations of our approach stem from uncertainties in several parameters, including the kick angles associated with the two kick processes, as well as the masses and orbital parameters of the pre-AIC progenitor. These uncertainties contribute to broad ranges in the derived kick velocities. Future observations could help refine these parameters and improve the constraints on kick velocities and the initial spin period. Such advancements would not only validate the formation history of GX 1$+$4 but also enhance our understanding of kick physics across various compact object formation channels, particularly the relative contributions of natal kicks versus delayed rocket kicks in binary evolution.

\begin{acknowledgements}
We are grateful to the referee for valuable comments that helped improve the manuscript. We thank Ze-Cheng Zou and Jun Yao for helpful discussion. This work was supported by the National Key Research and Development Program of China (2021YFA0718500) and the Natural Science Foundation of
China under grant Nos. 12041301, 12121003 and 123B2045.
\end{acknowledgements}

% \bibliography{ms.bib}
\bibliographystyle{aasjournalv7}

\end{CJK*}
\end{document}